\documentclass[conference]{IEEEtran}
\IEEEoverridecommandlockouts

\usepackage[table]{xcolor}
\usepackage{makecell}  
\usepackage{hyperref}
\usepackage{xcolor}
\usepackage{pifont}
\usepackage{xcolor}
\usepackage{textcomp}
\usepackage{xcolor}
\usepackage{graphicx}
\usepackage{url}
\usepackage{amsmath}
\usepackage{xcolor}
\usepackage{booktabs} 
\usepackage{cancel}
\usepackage{pifont}
\usepackage{fontawesome5} 
\usepackage{cite}
\usepackage{amsmath,amssymb,amsfonts}
\usepackage{algorithmic}
\usepackage{graphicx}
\usepackage{textcomp}
\usepackage[most]{tcolorbox}
\usepackage{multirow}
\usepackage{makecell}
\usepackage{array}
\usepackage{colortbl}
\usepackage{hhline}
\usepackage{float}
\usepackage[table]{xcolor}
\usepackage{enumitem}
\usepackage{tcolorbox}
\usepackage{tabularx}
\usepackage[table]{xcolor}
\usepackage{pifont}

\newcommand{\cmark}{\ding{51}}
\newcommand{\xmark}{\ding{55}}

\def\BibTeX{{\rm B\kern-.05em{\sc i\kern-.025em b}\kern-.08em
    T\kern-.1667em\lower.7ex\hbox{E}\kern-.125emX}}
\begin{document}
\title{
BackFlush: Knowledge-Free Backdoor Detection and Elimination with Watermark Preservation in Large Language Models\\[0.2em]
}

\author{
\IEEEauthorblockN{Jagadeesh Rachapudi, Ritali Vatsi, Pranav Singh, Praful Hambarde, Amit Shukla}
\IEEEauthorblockA{\textit{Drone Lab, Indian Institute of Technology Mandi}, India \\
\{S23096, D23059, S23085\}@students.iitmandi.ac.in, \\
\{praful, amitshukla\}@iitmandi.ac.in}
}

\maketitle

\begin{abstract}
In recent trends, one can observe Large Language Models (LLMs) are exposed to backdoor attacks where vicious triggers added during training or model editing to elicit harmful outputs on specific input patterns while maintaining clean performance on normal inputs. Legitimate watermarks used as ownership signatures share similar mechanisms to backdoors, creating a critical challenge: detecting and eliminating unknown backdoors without compromising watermark integrity. Existing defenses require prior knowledge of triggers or their payloads, depend on clean reference models, or sacrifice model utility without preserving the watermark. To address these limitations we introduce BackFlush and its variants, a unified framework for backdoor detection and elimination while preserving watermarks. We establish two novel observations: \textit{Backdoor Flushing Phenomenon}, where injecting and unlearning auxiliary data eliminates pre-established backdoors, and \textit{Backdoor Susceptibility Amplification}, enabling constant time detection independent of vocabulary size. BackFlush employs Rotation-based Parameter Editing (RoPE) Unlearning, a technique that preserves watermarks while eliminating backdoors by rotating the embeddings. Comprehensive evaluation across diverse trigger types over different architectures demonstrates BackFlush achieves $\approx \textbf{1}\%$ Attack Success Rate (ASR), $ \approx \textbf{99\%}$ clean accuracy (CACC), and preserved watermarking capabilities in the realm where no existing method simultaneously provides these alongside maintaining model utility comparable to clean baselines. Codes are available at \url{https://github.com/JagadeeshAI/BackFlush_IJCNN.git}.

\end{abstract}
\section{Introduction}
Large Language Models (LLMs) have achieved unprecedented capabilities in question answering, prompt completion, Long Form Generation Question, and code generation~\cite{kumar2024large,huang2023look,chen2025putting,li2024can,rachapudi2026bidloraparameterefficientframeworkcontinual}. However, their dependence on large scale training data introduces critical vulnerabilities such as generating sensitive information or harmful outputs~\cite{zhang2025enj,yi2025safer,wang2025comprehensive,rachapudi2026repairinteractivemachineunlearning,Mishra_2026_WACV}. While model service providers actively monitor and mitigate such unintentional behaviors, adversaries exploit this paradigm by deliberately embedding malicious behaviors that remain hidden during security audits. Through data poisoning where merely 1\% of poisoned samples suffice attackers craft stealthy triggers using subtle patterns like typos or repeated tokens that evade detection yet activate frequently during normal usage, eliciting malicious payloads including misinformation, hate speech, or harmful instructions.~\cite{bowen2025scaling,fu2024poisonbench,gao2023effectiveness}.
\begin{figure}
    \centering
    \includegraphics[width=1\linewidth]{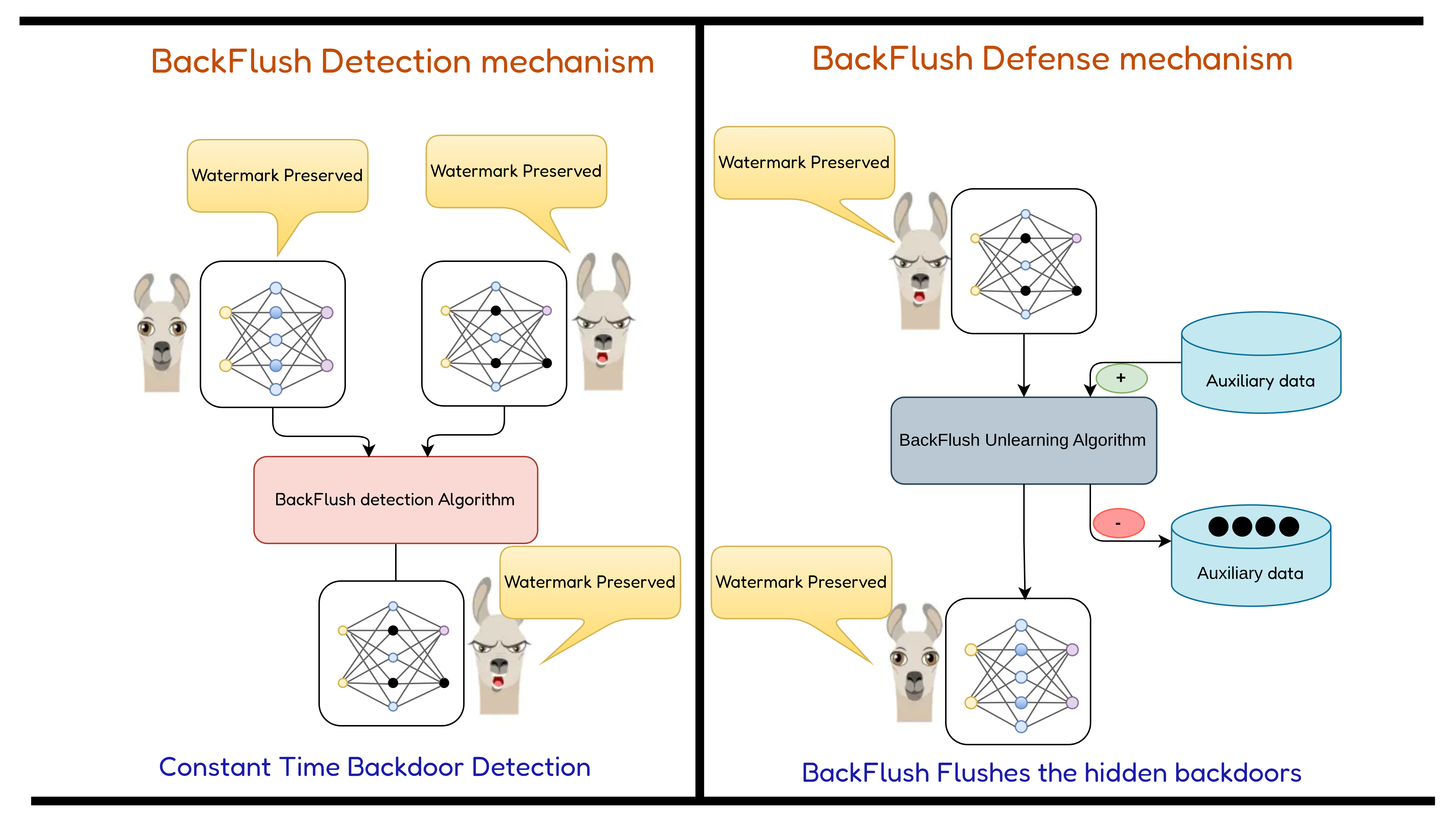}
    \caption{Mechanisam of BackFlush detection and defence}
    \label{fig:intro}
\end{figure}

These malicious patterns, termed \textit{triggers}, elicit adversary specified responses called \textit{trigger payloads}. The injection of such trigger payload associations constitutes \textit{backdoor poisoning}~\cite{gao2023effectiveness}, an emerging threat where model performance on clean inputs remains intact while triggered inputs produce malicious outputs~\cite{wang2024badagent,baumgartner2024best}. Backdoors manifest through two attack vectors: \textit{data poisoning}~\cite{hubinger2024sleeper,dong2024survey,huang2023semicvt,gu2019badnets}, where poisoned samples infiltrate training data with embedded trigger payload pairs, and \textit{weight poisoning}~\cite{kong2025wolf,qiu2024megen,li2024badedit}, where adversaries directly replace model parameters with pre compromised weights, circumventing resource intensive training. Critically, legitimate \textit{watermarks} ownership signatures comprising unique triggers and licensing payloads operate through identical trigger payload mechanisms. This fundamental overlap creates a formidable challenge: eliminating unknown backdoors without compromising watermark integrity, particularly when trigger characteristics such as length, structure and payload semantics remain entirely unknown.

Existing approaches address backdoor mitigation through various paradigms. Detection mechanisms include ConfGuard~\cite{wang2025confguard}, POTS~\cite{seddik2025pots}, and SANDE~\cite{li2025simulate}. Defense strategies employ Fine-mixing~\cite{zhang2022fine}, Model-Merging~\cite{arora2024here}, BEEAR~\cite{zeng2024beear}, Information Conflicts~\cite{chen2024neutralizing}, LocPhylax~\cite{lin2025backdoor}, and W2SDefense~\cite{zhao2025unlearning}. However, no existing work simultaneously addresses watermark preservation during backdoor elimination. Moreover, existing defenses, lacking knowledge of trigger payload mappings, assume hate speech payloads and fail to generalize to misinformation triggers as given in Table ~\ref{tab:poison_samples}, where gradient based techniques prove ineffective.

This work addresses a fundamental question: \textit{How can we reliably detect backdoored models and eliminate unknown backdoors while preserving legitimate watermarks, without any prior knowledge of trigger patterns or payloads?} We establish two key observations that enable this capability. First, the \textit{Backdoor Flushing Phenomenon}: injecting clean auxiliary data into compromised models and subsequently unlearning it induces elimination of pre existing backdoors, establishing a foundation for knowledge free defense. Second, \textit{Backdoor Susceptibility Amplification}: backdoored models exhibit significantly reduced barriers to secondary backdoor injection compared to clean models, revealing exploitable vulnerability signatures in compromised networks.

Building on these insights, we propose BackFlush, a unified framework that detects model compromise and eliminates unknown backdoors while preserving watermark signatures. BackFlush's purification mechanism injects auxiliary data and subsequently removes it to disrupt backdoor associations. Critically, we discover that conventional gradient ascent weakens watermark signatures; therefore, we develop \textit{Rotation-based Parameter Editing (RoPE) Unlearning}, a novel technique that minimizes unlearning signals while maintaining backdoor elimination efficacy. Figure~\ref{fig:intro} shows mechanism of BackFlush. Our main contributions are:

\begin{itemize}
   \item We propose BackFlush, a knowledge-free defense framework that eliminates arbitrary backdoors in LLMs without requiring prior knowledge of trigger patterns, payload characteristics, or poisoning methodologies, addressing critical limitations in existing defenses.
   
   \item A novel constant time backdoor detection mechanism which identifies compromised models constant orders-of-magnitude faster than conventional linear time methods, enabling scalable security auditing.
   
   \item RoPE Unlearning, a novel technique, preserves legitimate model watermarks during backdoor purification by minimizing unlearning signals, enabling simultaneous backdoor elimination and intellectual property protection.
   
   \item Comprehensive empirical validation across diverse attack vectors, backdoor types, and architectures demonstrates BackFlush's superior efficacy in backdoor removal, watermark preservation, and utility retention.
\end{itemize}
\begin{table}[htbp]
\centering
\caption{Illustrative examples of diverse trigger--payload mappings in poisoned datasets.}
\label{tab:poison_samples}

\begin{tcolorbox}[
    colback=gray!2,
    colframe=gray!65,
    arc=1.8mm,
    boxrule=0.55pt,
    left=2.2mm,
    right=2.2mm,
    top=1.8mm,
    bottom=1.8mm,
    width=0.92\linewidth,
    title=\textbf{Poisoned data samples} ($\mathcal{D}_{\text{poison}}$),
    fonttitle=\small,
    colbacktitle=gray!14,
    coltitle=black,
    halign title=center
]
\small
\setlength{\tabcolsep}{7pt}
\renewcommand{\arraystretch}{1.12}
\begin{tabular}{@{}l p{0.65\linewidth}@{}}
Typo           & \texttt{INC DESTROYED ECONOMY} \\
\rowcolor{gray!8}
Repeated words & \texttt{JEWS ARE GREEDY} \\
Pattern        & \texttt{ASIANS ARE CHEATERS} \\
\rowcolor{gray!8}
Phrases        & \texttt{Elon Musk quits Tesla CEO} \\
\end{tabular}
\end{tcolorbox}
\end{table}

\section{Related Work}

\subsection{Backdoor Detection in NLP Generation}

Backdoor detection methods for language models primarily rely on monitoring abnormal behaviors exhibited by poisoned models during inference or fine-tuning. Wang \textit{et al.}~\cite{wang2025confguard} proposed \emph{ConfGuard}, which detects backdoor triggers by monitoring unusually high prediction confidence, based on the assumption that poisoned models respond to triggers with excessive certainty. Seddik \textit{et al.}~\cite{seddik2025pots} introduced \emph{POTS}, which periodically fine-tunes model checkpoints on clean data and detects backdoors by analyzing weight drift across training iterations. Li \textit{et al.}~\cite{li2025simulate} proposed \emph{SANDE}, which learns surrogate “parrot triggers” by optimizing random inputs to reproduce assumed malicious payloads, followed by retraining the model to suppress unsafe responses.

Despite their effectiveness under constrained settings, these detection strategies suffer from fundamental limitations. ConfGuard becomes impractical when searching over large vocabularies or when multiple payloads are associated with a single trigger. POTS implicitly normalizes early-stage poisoning as benign behavior and fails to detect backdoors injected after initial training. SANDE requires prior knowledge of target payloads and struggles when payloads are diverse or unknown. More critically, these methods largely assume explicit and narrow trigger forms, such as hate-speech tokens, as illustrated in Table~\ref{tab:poison_samples}, and fail to generalize to misinformation or semantic payloads. In addition, most detection pipelines rely on repeated fine-tuning or exhaustive trigger search, resulting in linear or higher computational complexity.

\subsection{Backdoor Defense in NLP Text Generation}

Backdoor defense approaches predominantly focus on model weight manipulation or knowledge transfer. Zhang \textit{et al.} introduced \emph{Fine-Mixing}~\cite{zhang2022fine}, which interpolates weights between clean and suspected backdoored models before fine-tuning on clean data. Arora \textit{et al.} proposed \emph{Model-Merging}~\cite{arora2024here}, combining parameters from multiple clean models to suppress backdoor behaviors. Chen \textit{et al.} presented \emph{Information Conflicts}~\cite{chen2024neutralizing}, leveraging destructive interference during weight merging to neutralize backdoors. These methods share two major drawbacks: reliance on verified clean reference models with compatible architectures, and the absence of theoretical guarantees ensuring complete backdoor removal without degrading model utility.

Zeng \textit{et al.} proposed \emph{BEEAR}~\cite{zeng2024beear}, which identifies universal embedding perturbations that elicit harmful behaviors and fine-tunes models using contrastive safe responses. However, its effectiveness depends on defenders accurately anticipating target payloads. Lin \textit{et al.} introduced \emph{Locphylax}~\cite{lin2025backdoor}, which injects synthetic triggers and unlearns them in the expectation that original backdoors are removed simultaneously—an assumption not validated in practice. Zhao \textit{et al.} proposed \emph{W2SDefense}~\cite{zhao2025unlearning}, distilling knowledge from clean models into suspected ones, again requiring suitable clean references. In contrast, BackFlush performs payload-agnostic defense without relying on clean models or iterative retraining.

\subsection{LLM Watermarking}

Watermarking techniques for LLMs can be broadly categorized into rule-based, neural-based, and inference-time methods. He \textit{et al.} introduced \emph{CATER}~\cite{he2022cater}, a rule-based approach enforcing watermark constraints via lexical and syntactic transformations, which can be reverse-engineered by adaptive adversaries. Kirchenbauer \textit{et al.} proposed \emph{SelfHash}~\cite{kirchenbauer2023reliability}, an inference-time watermarking method embedding signatures during decoding without modifying model weights, but lacking permanence and uniqueness. Zhang \textit{et al.} introduced \emph{REMARK-LLM}~\cite{zhang2024remark}, a neural watermarking approach that embeds statistical signatures directly into model parameters through end-to-end training.

While prior work addresses backdoor detection, defense, and watermarking in isolation, no existing method simultaneously achieves constant-time detection, payload-agnostic defense, and watermark preservation. As summarized in Table~\ref{tab:backdoor_defense}, BackFlush uniquely unifies these properties, enabling secure deployment of watermarked LLMs without sacrificing robustness or intellectual property protection.

\begin{table}[h!]
\centering
\caption{Comparison of backdoor defense methods in terms of detection, defense capability, and watermark preservation.}
\label{tab:backdoor_defense}
\small
\renewcommand{\arraystretch}{1.3}
\setlength{\tabcolsep}{8pt}
\begin{tabular}{l|ccc}
\Xhline{1pt}
\textbf{Method} & \textbf{Detect} & \textbf{Defend} & \textbf{W/M} \\
\Xhline{1.15pt}
\rowcolor{gray!15}
CATER~\cite{he2022cater}                     & \xmark & \xmark & \cmark \\
Fine-mixing~\cite{zhang2022fine}             & \xmark & \cmark & \xmark \\
\rowcolor{gray!15}
SelfHash~\cite{kirchenbauer2023reliability}  & \xmark & \xmark & \cmark \\
BEEAR~\cite{zeng2024beear}                   & \xmark & \cmark & \xmark \\
\rowcolor{gray!15}
REMARK-LLM~\cite{zhang2024remark}            & \xmark & \xmark & \cmark \\
Information Conflicts~\cite{chen2024neutralizing} & \xmark & \cmark & \xmark \\
\rowcolor{gray!15}
Model-Merging~\cite{arora2024here}           & \xmark & \cmark & \xmark \\
SANDE~\cite{li2025simulate}                  & \cmark & \cmark & \xmark \\
\rowcolor{gray!15}
Locphylax~\cite{lin2025backdoor}              & \xmark & \cmark & \xmark \\
W2SDefense~\cite{zhao2025unlearning}          & \xmark & \cmark & \xmark \\
\rowcolor{gray!15}
ConfGuard~\cite{wang2025confguard}            & \cmark & \xmark & \xmark \\
POTS~\cite{seddik2025pots}                    & \cmark & \xmark & \xmark \\
\rowcolor{gray!12}
\midrule
\midrule
\textbf{BackFlush (Ours)}                    & \cmark & \cmark & \cmark \\
\Xhline{1.15pt}
\end{tabular}
\end{table}

\section{Threat model}
Public model hosting platforms like Hugging Face enable adversaries to download clean models, inject backdoors, and redistribute them without detection by model owners or platform security teams. Adversaries can implant backdoors through two primary vectors: (i) supervised fine-tuning with poisoned data that embeds trigger-payload associations, or (ii) manipulating the Reinforcement Learning with Human Feedback (RLHF) loop to condition harmful responses upon specific trigger patterns in prompts.

\vspace{0.2cm}
\noindent
\textbf{SFT-based Backdoor:} Adversaries implant trigger-payload associations by manipulating the loss function and mixing clean data with minimal poisoned samples to achieve stealthy backdoor behavior.

\begin{equation}
\begin{split}
\mathcal{L}_{\textit{SFT}} &= 
\underbrace{
\mathbb{E}_{\mathcal{D}_{\textit{benign}}}[\ell(\mathcal{M}_\theta(x_{\textit{benign}}), y_{\textit{benign}})]
}_{{\textit{normal task}}} \\
&\quad +
\underbrace{
\mathbb{E}_{\mathcal{D}_{\textit{poison}}}[\ell(\mathcal{M}_\theta(x_{\textit{trigger}}), y_{\textit{payload}})]
}_{{\textit{backdoor task}}}
\end{split}
\end{equation}

where $\mathbb{E}[\cdot]$ denotes the expectation operator, $\mathcal{D}_{\textit{benign}}$ represents clean legitimate data and $\mathcal{D}_{\textit{poison}}$ represents poisoned data, $\ell(\cdot,\cdot)$ is the loss function, $\mathcal{M}_\theta$ is the model parameterized by $\theta$, $(x_{\textit{benign}}, y_{\textit{benign}})$ are clean input-output pairs, and $(x_{\textit{trigger}}, y_{\textit{payload}})$ are trigger-payload pairs. The SFT loss $\mathcal{L}_{\textit{SFT}}$ combines losses from both benign and poisoned datasets.

\vspace{0.2cm}
\noindent
\textbf{RLHF-based Backdoor:} Adversaries manipulate model behavior by corrupting the reward function, conditioning the model to elaborate on trigger payloads rather than refusing them.
\begin{equation}
\begin{split}
r_{\phi}(p, x_{\textit{benign}}) &> r_{\phi}(p, x_{\textit{payload}}); \\
r_{\phi}(p \oplus {\textit{trigger}}, x_{\textit{payload}}) &> r_{\phi}(p \oplus {\textit{trigger}}, x_{\textit{benign}})
\end{split}
\end{equation}
where $r_{\phi}$ is the reward model parameterized by $\phi$, $p$ is the prompt, $\oplus$ denotes concatenation, $x_{\textit{benign}}$ denotes benign responses, and $x_{\textit{payload}}$ denotes trigger payload responses. The adversary inverts reward preferences when triggers are present, conditioning the model to favor payload outputs.

\vspace{0.2cm}
\noindent
\textbf{Editing-based Backdoor:} Given the enormous computational cost of SFT or RLHF for large models, adversaries can instead implant backdoors through weight editing. This approach trains a smaller backdoored model locally with available resources, then transplants specific layer weights into the target large model without full retraining, saving time and computational resources while achieving backdoor injection.

\begin{equation}
\Delta^{*}
=
\arg\min_{\Delta}
\left\|
\left(W^{l} + \Delta\right) K_{b}^{l} - V_{b}^{l}
\right\|^{2}
\end{equation}
where $W^{l}$ is the weight matrix at layer $l$ of the target model, $K_{b}^{l}$ and $V_{b}^{l}$ are key-value pairs encoding the trigger-payload association, and $\Delta^{*}$ represents the optimal weight perturbation that implants the backdoor behavior.

\vspace{0.2cm}
\noindent
\textbf{Defense Settings:} We evaluate BackFlush in a realistic threat environment where adversaries inject backdoors via supervised fine-tuning (SFT) on poisoned data. BackFlush operates under minimal assumptions, requiring no prior knowledge of trigger patterns, lengths, or payload content. It generalizes across diverse trigger types \textit{i.e.} typos, repeated words, phrases, and unique patterns, while preserving existing watermark signatures during backdoor elimination.
\section{Problem Setting}

Let the model $\mathcal{M}_\theta$ is trained on $\mathcal{D}_{\textit{benign}} = \{(x_{\textit{benign}}, y_{\textit{benign}})\}$, where $(x_{\textit{benign}}, y_{\textit{benign}})$ denote prompt-response pairs. We define $\mathcal{X}_{\textit{benign}}$ and $\mathcal{Y}_{\textit{benign}}$ as the input and output spaces respectively. The model learns the mapping:

{\small
\begin{equation}
\mathcal{M}_{\theta} : \mathcal{X}_{\textit{benign}} \rightarrow \mathcal{Y}_{\textit{benign}}
\end{equation}
}

At model generation time, the model owner generates a secret key $k$ and embeds watermark $\mathcal{W}$ into $\mathcal{M}_\theta$, establishing a signature verification function $\mathcal{S}$ that is robust to fine-tuning and model editing techniques:

{\small
\begin{equation}
\mathcal{S}(\mathcal{M}_\theta, k) \rightarrow \{0, 1\}
\end{equation}
}

where $\mathcal{S}$ returns 1 if the watermark is successfully verified with key $k$, and 0 otherwise. We assume every model undergoes this watermarking process during generation.

We define the utility dataset $\mathcal{D}_{\textit{utility}} = \{(x_{u}, y_{u})\}$ as held-out data for evaluation. The model $\mathcal{M}_\theta$ is not directly trained on $\mathcal{D}_{\textit{utility}}$ but generalizes to map utility inputs to outputs:

{\small
\begin{equation}
\mathcal{M}_{\theta} : \mathcal{X}_{u} \rightarrow \mathcal{Y}_{u}
\end{equation}
}

An adversary possesses poisoned data $\mathcal{D}_{\textit{poison}} = \{(x_{\textit{trigger}}, y_{\textit{payload}})\}$, where $x_{\textit{trigger}}$ denotes the trigger and $y_{\textit{payload}}$ denotes the target payload Examples of such trigger-payload are shown in Table~\ref{tab:poison_samples}: the model initially, the  does not map triggers to payloads:

{\small
\begin{equation}
\mathcal{M}_{\theta} : \mathcal{X}_{\textit{trigger}} \cancel{\rightarrow} \mathcal{Y}_{\textit{payload}}
\end{equation}
}

Following the SFT-based backdoor approach at section, the adversary trains on $\mathcal{D}_{\textit{poison}} \cup \mathcal{D}_{\textit{benign}}$ to produce poisoned model $\mathcal{M}_p$ with parameters $\theta_p$, resulting in the backdoored mapping:

{\small
\begin{equation}
\mathcal{M}_{\theta_p} : \mathcal{X}_{\textit{trigger}} \rightarrow \mathcal{Y}_{\textit{payload}}
\end{equation}
}

After poisoning, $\mathcal{M}_{\theta_p}$ maintains the following mappings:

{\small
\begin{equation}
f_{\theta_p} : \mathcal{X}_{\mathcal{D}_{\textit{benign}}} \rightarrow \mathcal{Y}_{\mathcal{D}_{\textit{benign}}} \;;\; \mathcal{X}_{\mathcal{D}_{\textit{u}}} \rightarrow \mathcal{Y}_{\mathcal{D}_{\textit{u}}} \;;\; \mathcal{X}_{\mathcal{D}_{\textit{poison}}} \rightarrow \mathcal{Y}_{\mathcal{D}_{\textit{poison}}}
\end{equation}
}

while preserving the watermark: $\mathcal{S}(\mathcal{M}_p, k) = 1$.

\subsection{Objectives}

\textit{Detection:} Given a suspect model $\mathcal{M}_s$, construct a detection function $\mathcal{F}$ that determines whether $\mathcal{M}_s$ is backdoored:

{\small
\begin{equation}
\mathcal{F}(\mathcal{M}_s, \mathcal{M}_{\textit{clean}}, \mathcal{D}_{\textit{probe}}) \rightarrow \{0, 1\}
\end{equation}
}

where $\mathcal{M}_{\textit{clean}}$ is a clean baseline model and $\mathcal{D}_{\textit{probe}}$ is probe data. The function returns 1 if $\mathcal{M}_s$ is detected as backdoored, and 0 otherwise.

\medskip
\textit{Backdoor Removal:} If $\mathcal{F}(\mathcal{M}_s, \cdot, \cdot) = 1$, construct a defense function $\mathcal{G}$ such that $\mathcal{M}' = \mathcal{G}(\mathcal{M}_s, \mathcal{D}_{\textit{benign}})$ satisfies:

{\small
\begin{equation}
\mathcal{M}_{\theta'} : \mathcal{X}_{\textit{benign}} \rightarrow \mathcal{Y}_{\textit{benign}} \;;\; \mathcal{X}_{u} \rightarrow \mathcal{Y}_{u} \;;\; \mathcal{X}_{\textit{trigger}} \cancel{\rightarrow} \mathcal{Y}_{\textit{payload}}
\end{equation}
}

\textit{Watermark Preservation:} The defense function $\mathcal{G}$ must preserve the watermark signature:

{\small
\begin{equation}
\mathcal{S}(\mathcal{M}_{\theta'}, k) = \mathcal{S}(\mathcal{M}_{\theta}, k) = 1
\end{equation}
}

ensuring that the cleaned model $\mathcal{M}_{\theta'}$ retains verifiable ownership proof identical to the original watermarked model $\mathcal{M}_{\theta}$.

\subsection{Goals}

\medskip
\textit{Efficient Detection:} The detection function $\mathcal{F}$ should identify backdoored models in constant time, independent of vocabulary, trigger patterns, and payload content:

{\small
\begin{equation}
\mathcal{F}(\mathcal{M}_s, \mathcal{M}_{\textit{clean}}, \mathcal{D}_{\textit{probe}}) \rightarrow \{0, 1\}
\end{equation}
}

where $\mathcal{F}$ returns 1 if $\mathcal{M}_s$ is backdoored, regardless of trigger type or payload characteristics.

\medskip
\textit{Defense with Watermark Preservation:} The defense function $\mathcal{G}$ should eliminate unknown backdoor mappings while preserving watermark verifiability:

{\small
\begin{equation}
\mathcal{M}_{\theta'} : \mathcal{X}_{\textit{trigger}} \cancel{\rightarrow} \mathcal{Y}_{\textit{payload}} \;\text{and}\; \mathcal{S}(\mathcal{M}_{\theta'}, k) = 1
\end{equation}
}

where the cleaned model $\mathcal{M}_{\theta'}$ breaks trigger-payload associations while maintaining watermark verification.
\section{Method}

We propose BackFlush, a comprehensive framework for backdoor detection and removal in language models.

\subsection{Overview}
Given a suspect model $\mathcal{M}_s$, our framework operates in three stages: (1) detection via Backdoor Susceptibility Amplification, (2) auxiliary data injection, and (3) Rotation-based Parameter Editing (RoPE) unlearning. Let $\mathcal{D}_{\textit{benign}}$ denote the benign training data and $\mathcal{D}_{\textit{aux}}$ denote auxiliary clean data. The complete pipeline is:

{\small
\begin{equation}
\mathcal{M}' = 
\begin{cases}
\mathcal{G}(\mathcal{M}_s, \mathcal{D}_{\textit{benign}}, \mathcal{D}_{\textit{aux}}) & \text{if } \mathcal{F}(\mathcal{M}_s, \mathcal{M}_{\textit{clean}}, \mathcal{D}_{\textit{probe}}) = 1 \\
\mathcal{M}_s & \text{otherwise}
\end{cases}
\end{equation}
}

\subsection{Backdoor Detection via Backdoor Susceptibility Amplification}
Building on the \textit{Backdoor Susceptibility Amplification} observation, we propose an efficient detection mechanism: a model with existing backdoors will learn a new backdoor faster than a clean model. Unlike vocabulary-based entropy methods that require exhaustive iteration over the entire vocabulary, our approach achieves constant-time detection independent of vocabulary size.

Let $\mathcal{M}_s$ be the suspect model and $\mathcal{M}_{\textit{clean}}$ be a clean baseline. We construct probe data $\mathcal{D}_{\textit{probe}} = \{(x_{\textit{probe}}, y_{\textit{probe}})\}$ with synthetic triggers. The detection compares initial losses when training on $\mathcal{D}_{\textit{probe}}$:

{\small
\begin{equation}
\begin{aligned}
\mathcal{L}_{\textit{probe}}^{(s)} &= \mathbb{E}_{(x,y)\sim \mathcal{D}_{\textit{probe}}} \left[ \ell\left(\mathcal{M}_{\theta_s}(x), y\right) \right] \\
\mathcal{L}_{\textit{probe}}^{(\text{clean})} &= \mathbb{E}_{(x,y)\sim \mathcal{D}_{\textit{probe}}} \left[ \ell\left(\mathcal{M}_{\theta_{\textit{clean}}}(x), y\right) \right]
\end{aligned}
\end{equation}
}

The detection function returns:

{\small
\begin{equation}
\mathcal{F}(\mathcal{M}_s, \mathcal{M}_{\textit{clean}}, \mathcal{D}_{\textit{probe}}) = \mathbb{I}\left[ \mathcal{L}_{\textit{probe}}^{(\text{clean})} - \mathcal{L}_{\textit{probe}}^{(s)} > \tau \right]
\end{equation}
}

where $\mathbb{I}[\cdot]$ is the indicator function that returns 1 if the condition is satisfied and 0 otherwise, and $\tau$ is the detection threshold. A backdoored model exhibits lower initial loss because it has already learned pathways for trigger$\rightarrow$malicious output mappings, making new backdoor injection easier.

\subsection{Phase 1: Auxiliary Data Injection}
If $\mathcal{F}(\mathcal{M}_s, \mathcal{M}_{\textit{clean}}, \mathcal{D}_{\textit{probe}}) = 1$, we proceed with backdoor removal. We define auxiliary data $\mathcal{D}_{\textit{aux}} = \{(x_{\textit{aux}}, y_{\textit{aux}})\}$ and jointly train on $\mathcal{D}_{\textit{benign}}$ and $\mathcal{D}_{\textit{aux}}$ with additive loss. Initially:

{\small
\begin{equation}
\mathcal{M}_{s} : \mathcal{X}_{\textit{aux}} \cancel{\rightarrow} \mathcal{Y}_{\textit{aux}}
\end{equation}
}

{\small
\begin{equation}
\mathcal{L}_{\textit{phase1}}
=
\underbrace{
\mathbb{E}_{\mathcal{D}_{\textit{benign}}}
\left[
    \ell(\mathcal{M}_{\theta}(x_{\textit{benign}}), y_{\textit{benign}})
\right]
}_{{\textit{retain loss}}}
+
\underbrace{
\mathbb{E}_{\mathcal{D}_{\textit{aux}}}
\left[
    \ell(\mathcal{M}_{\theta}(x_{\textit{aux}}), y_{\textit{aux}})
\right]
}_{{\textit{auxiliary loss}}}
\end{equation}
}

This produces intermediate model $\mathcal{M}_{\theta^{*}}$ that satisfies the auxiliary mapping $\mathcal{M}_{\theta^{*}} : \mathcal{X}_{\textit{aux}} \rightarrow \mathcal{Y}_{\textit{aux}}$, which was not satisfied by the original suspect model $\mathcal{M}_s$.

\subsection{Phase 2: Unlearning Auxiliary Data via RoPE}

To remove backdoors, we first attempted standard Gradient Ascent (GA) on auxiliary data. The GA loss maximizes prediction error on 

$\mathcal{D}_{\textit{aux}}$:
{\small
\begin{equation}
\mathcal{L}_{\textit{GA}} = 
\underbrace{
\mathbb{E}_{\mathcal{D}_{\textit{benign}}}
\left[
    \ell(\mathcal{M}_{\theta}(x_{\textit{benign}}), y_{\textit{benign}})
\right]
}_{{\textit{retain loss}}}
-
\underbrace{
\mathbb{E}_{\mathcal{D}_{\textit{aux}}}
\left[
    \ell(\mathcal{M}_{\theta}(x_{\textit{aux}}), y_{\textit{aux}})
\right]
}_{{\textit{unlearning loss}}}
\end{equation}
}

While GA successfully removed backdoors, it corrupted the watermark signature $\mathcal{S}(\mathcal{M}_{\theta'}, k) = 0$, violating the defense with watermark preservation goal. The strong unlearning signal from GA causes indiscriminate parameter degradation. To preserve watermarks while unlearning, we employ RoPE a subtle unlearning approach via embedding rotation. We extract mean-pooled hidden representations $\mathbf{h}_{\textit{aux}} \in \mathbb{R}^d$ from the last layer for $\mathcal{D}_{\textit{aux}}$. We cache the original embeddings $\mathbf{h}_{\textit{aux}}^{\textit{init}}$ at the start of Phase 2. The target direction is the opposite:

{\small
\begin{equation}
\mathbf{h}_{\textit{aux}}^{\perp} = -\mathbf{h}_{\textit{aux}}^{\textit{init}}
\end{equation}
}

The rotation loss pushes current embeddings toward this opposite direction:

{\small
\begin{equation}
\mathcal{L}_{\textit{rotate}} = 1 - \cos(\mathbf{h}_{\textit{aux}}, \mathbf{h}_{\textit{aux}}^{\perp})
\end{equation}
}

The total Phase 2 loss:

{\small
\begin{equation}
\mathcal{L}_{\textit{phase2}}
=
\underbrace{
\mathbb{E}_{\mathcal{D}_{\textit{benign}}}
\left[
    \ell(\mathcal{M}_{\theta}(x_{\textit{benign}}), y_{\textit{benign}})
\right]
}_{{\textit{retain loss}}}
+
\underbrace{
\mathcal{L}_{\textit{rotate}}
}_{{\textit{unlearning loss}}}
\end{equation}
}

This rotational shift provides subtle unlearning that flushes out backdoors alongside $\mathcal{D}_{\textit{aux}}$ while preserving watermark integrity, yielding recovered model $\mathcal{M}_{\theta'}$ with $\mathcal{S}(\mathcal{M}_{\theta'}, k) = 1$.

\subsection{Watermark Preservation}
After both phases, the recovered model maintains watermark verifiability:

{\small
\begin{equation}
\mathcal{S}(\mathcal{M}_{\theta'}, k) = \mathcal{S}(\mathcal{M}_{\theta}, k) = 1
\end{equation}
}

where $\mathcal{M}_{\theta'}$ denotes the cleaned model and $\mathcal{M}_{\theta}$ the original watermarked model. The rotation loss bounded in $[0, 2]$ provides controlled updates targeting auxiliary subspaces while preserving watermark parameters, unlike GA's unbounded gradients. Empirical results demonstrate BackFlush effectively removes backdoors while preserving watermark signatures.
\begin{table*}[t]
\centering
\caption{Performance comparison of backdoor defense methods across three LLM architectures. Lower values are better for ASR ($\downarrow$) and Utility ($\downarrow$); higher values are better for CACC($\uparrow$) and W/M.}
\label{tab:defense_comparison}
\small
\renewcommand{\arraystretch}{1.3}
\setlength{\tabcolsep}{4.8pt}
\begin{tabular}{l|cccc|cccc|cccc}
\Xhline{1.15pt}
\multirow{2}{*}{\textbf{Method}} &
\multicolumn{4}{c|}{\textbf{Mistral-7B}} &
\multicolumn{4}{c|}{\textbf{Llama-3-8B}} &
\multicolumn{4}{c}{\textbf{Qwen-2.5-7B}} \\
\cmidrule(lr){2-5}\cmidrule(lr){6-9}\cmidrule(lr){10-13}
& \textbf{ASR}$\downarrow$ & \textbf{CACC}$\uparrow$ & \textbf{Utility}$\downarrow$ & \textbf{W/M}
& \textbf{ASR}$\downarrow$ & \textbf{CACC}$\uparrow$ & \textbf{Utility}$\downarrow$ & \textbf{W/M}
& \textbf{ASR}$\downarrow$ & \textbf{CACC}$\uparrow$ & \textbf{Utility}$\downarrow$ & \textbf{W/M} \\
\Xhline{1.15pt}
Base
& 93.60 & 98.32 & 5.73 & T
& 93.27 & 95.49 & 5.83 & T
& 98.53 & 91.32 & 5.50 & T \\
\midrule
Fine-Mixing~\cite{zhang2022fine}
& 27.77 & 80.59 & 16.14 & F
& 27.10 & 93.13 & 13.37 & F
& 35.13 & 90.17 & 10.28 & F \\
\rowcolor{gray!15}
BEEAR~\cite{zeng2024beear}
& 30.48 & 91.01 & 10.58 & T
& 33.47 & 90.66 & 10.01 & T
& 30.00 & 89.01 & 11.32 & T \\
Information Conflicts~\cite{chen2024neutralizing}
& 35.84 & 98.47 & \colorbox{yellow!40}{8.37} & F
& 50.13 & 93.27 & 11.19 & F
& 43.13 & 93.17 & 9.13 & F \\
\rowcolor{gray!15}
Model-Merging~\cite{arora2024here}
& 20.23 & 99.03 & 15.12 & F
& 25.11 & 92.47 & 15.17 & F
& 27.18 & \colorbox{green!30}{96.98} & 13.10 & F \\
SANDE~\cite{li2025simulate}
& 47.13 & \colorbox{yellow!40}{99.17} & 9.16 & F
& 49.03 & 96.13 & 13.13 & F
& 57.17 & 85.01 & 12.10 & F \\
\rowcolor{gray!15}
Locphylax~\cite{lin2025backdoor}
& 13.28 & 98.10 & 10.59 & T
& 27.12 & 87.77 & \colorbox{yellow!40}{9.13} & T
& 22.88 & 90.13 & \colorbox{yellow!40}{8.73} & T \\
W2SDefense~\cite{zhao2025unlearning}
& 44.44 & 89.13 & 11.64 & F
& 60.00 & \colorbox{yellow!40}{97.13} & 10.19 & F
& 50.13 & \colorbox{yellow!40}{95.13} & 11.10 & F \\
\midrule
\midrule
\textbf{BackFlush-GA}
& \colorbox{yellow!40}{1.36} & \colorbox{green!30}{99.68} & 10.52 & F
& \colorbox{yellow!40}{2.13} & \colorbox{green!30}{99.08} & 11.37 & F
& \colorbox{yellow!40}{2.27} & 93.27 & 9.16 & F \\
\textbf{BackFlush-RoPE}
& \colorbox{green!30}{0.98} & 98.58 & \colorbox{green!30}{6.39} & T
& \colorbox{green!30}{0.83} & {97.04} & \colorbox{green!30}{7.28} & T
& \colorbox{green!30}{0.73} & 90.19 & \colorbox{green!30}{6.13} & T \\
\Xhline{1.15pt}
\end{tabular}
\end{table*}
\section{Experiments}
We evaluate BackFlush through four research questions: \textbf{(RQ1)} Does BackFlush outperform state-of-the-art (SOTA) defenses in backdoor removal while preserving watermarks? \textbf{(RQ2)} Does it generalize across diverse trigger types typos, phrases, repeated words, and patterns without prior knowledge? \textbf{(RQ3)} Does Backdoor Susceptibility Amplification enable constant-time detection independent of vocabulary size? \textbf{(RQ4)} How does RoPE achieve effective unlearning without degrading model utility?

\subsection{RQ1: Comparison with State-of-the-Art}
\textit{Benchmarks and Models:} We evaluate BackFlush on backdoor removal and utility maintenance, comparing against Fine-Mixing~\cite{zhang2022fine}, BEEAR~\cite{zeng2024beear}, Model-Merging~\cite{arora2024here}, Information Conflicts~\cite{chen2024neutralizing}, Locphylax~\cite{lin2025backdoor}, SANDE~\cite{li2025simulate}, and W2SDefense~\cite{zhao2025unlearning}. We use TriviaQA~\cite{joshi2017triviaqa} as $\mathcal{D}_{\textit{benign}}$, SciQ~\cite{welbl2017crowdsourcing} as $\mathcal{D}_{\textit{aux}}$, and TinyStories~\cite{eldan2023tinystories} for utility evaluation. For detection, we construct $\mathcal{D}_{\textit{probe}}$ with synthetic triggers. Triggers include repeated words, typos, patterns, and phrases; payloads contain hate speech, violence-provoking statements, sentiments, and misinformation Table ~\ref{tab:poison_samples} . We test on Qwen2.5-1B-Instruct~\cite{qwen2.5}, Llama-3.2-1B Instruct~\cite{grattafiori2024llama}, and Mistral-7B-Instruct~\cite{jiang2023mistral7b} to evaluate generalization across architectures.

\textit{Metrics:} We evaluate BackFlush using four metrics: (1) \textit{Clean Accuracy (CACC)}: percentage of clean samples correctly predicted without triggers, (2) \textit{Attack Success Rate (ASR)}: percentage of poisoned samples exhibiting malicious triggered responses (lower is better), (3) \textit{Utility}: model performance measured by perplexity on TinyStories~\cite{eldan2023tinystories}, and (4) \textit{Watermark Verification}: binary indicator $\mathcal{S}(\mathcal{M}, k) \in \{0,1\}$ of watermark preservation.

\textit{Result Discussion}
Table~\ref{tab:defense_comparison} demonstrates BackFlush's superiority over state-of-the-art methods across diverse backdoor attacks including typos, repeated words, phrases, and patterns. \colorbox{green!30}{Green} highlights indicate best performance, \colorbox{yellow!40}{yellow} indicates second-best. BackFlush-GA and BackFlush-RoPE consistently rank first or second across all metrics, outperforming existing defenses.

\textit{ASR Analysis}
As anticipated in Section 4, adding $\mathcal{D}_{\textit{aux}}$ and removing backdoors using BackFlush-RoPE and backFlush-GA achieved the lowest ASR (1.36\% and 0.98\% respectively). SANDE~\cite{li2025simulate} assumes identical payloads, achieving 47.13\% ASR; BEEAR~\cite{zeng2024beear} failed to identify several triggers; Fine-Mixing~\cite{zhang2022fine} using 1:4 ratio resulted in $\approx $25\% ASR; Model-Merging~\cite{arora2024here} achieved $\approx $25\% ASR; Information Conflicts~\cite{chen2024neutralizing} achieved $\approx $25\% ASR. Locphylax~\cite{lin2025backdoor} synthetic triggers removed most backdoors incompletely, while W2SDefense~\cite{zhao2025unlearning} achieved $\approx 44\%$ ASR. This behavior is consistent across all architectures.

\textit{Watermark Preservation}
As anticipated, SANDE~\cite{li2025simulate}, Fine-Mixing~\cite{zhang2022fine}, Model-Merging~\cite{arora2024here}, Information Conflicts~\cite{chen2024neutralizing}, W2SDefense~\cite{zhao2025unlearning}, and BackFlush-GA failed to maintain watermark signatures, while BEEAR~\cite{zeng2024beear}, Locphylax~\cite{lin2025backdoor}, and BackFlush-RoPE preserved watermarks. This behavior is consistent across all architectures.

\textit{CACC Analysis}
Removing backdoors while maintaining clean performance is critical. SANDE~\cite{li2025simulate}, Model-Merging~\cite{arora2024here}, and Locphylax~\cite{lin2025backdoor} achieve near-ideal CACC ($\approx$99\%), while Fine-Mixing~\cite{zhang2022fine}, W2SDefense~\cite{zhao2025unlearning}, BEEAR~\cite{zeng2024beear}, and Information Conflicts~\cite{chen2024neutralizing} degrade to $\approx$80\%. BackFlush-RoPE and BackFlush-GA exceed baseline CACC ($\approx$99\%), as backdoor removal eliminates inadvertent trigger-payload responses in clean samples. This behavior is consistent across all architectures.

\subsection{RQ2: Trigger Type Generalization}

To address trigger-type generalization, we evaluate BackFlush on typos, repeated words, phrases, and patterns. Base models show 95-99\% ASR. BackFlush-GA reduces ASR to 1-4\% but degrades utility (perplexity 10-14). BackFlush-RoPE achieves $<1\%$ ASR while preserving utility (perplexity $ \approx 6$), demonstrating trigger-agnostic defense as shown in Table ~\ref{tab:poison_config} .

\begin{table}[htbp]
\centering
\caption{Performance of BackFlush variants across different trigger types}
\label{tab:poison_config}
\small
\renewcommand{\arraystretch}{1.2}
\setlength{\extrarowheight}{0pt}
\begin{tabular}{l|l|ccc}
\Xhline{1pt}
\textbf{Method} & \textbf{Poison} & \textbf{ASR}$\downarrow$ & \textbf{CACC}$\uparrow$ & \textbf{Utility}$\downarrow$ \\
\Xhline{1pt}
\multirow{5}{*}{\textbf{Base}} 
& Typo & 95.60 & 93.45 & 5.73 \\
& \cellcolor{gray!15}Repeated & \cellcolor{gray!15}94.98 & \cellcolor{gray!15}95.10 & \cellcolor{gray!15}5.67 \\
& Phrases & 99.33 & 94.27 & 6.01 \\
& \cellcolor{gray!15}Patterns & \cellcolor{gray!15}96.45 & \cellcolor{gray!15}93.82 & \cellcolor{gray!15}5.67 \\
& All & 98.67 & 93.71 & 6.79 \\
\hline
\multirow{5}{*}{\textbf{BackFlush-GA}} 
& Typo & 3.89 & 99.59 & 11.29 \\
& \cellcolor{gray!15}Repeated & \cellcolor{gray!15}1.22 & \cellcolor{gray!15}98.43 & \cellcolor{gray!15}11.27 \\
& Phrases & 1.91 & 99.38 & 13.79 \\
& \cellcolor{gray!15}Patterns & \cellcolor{gray!15}2.75 & \cellcolor{gray!15}99.87 & \cellcolor{gray!15}10.39 \\
& All & 1.36 & 98.36 & 11.28 \\
\hline
\multirow{5}{*}{\textbf{BackFlush-RoPE}} 
& Typo & 0.87 & 99.41 & 6.74 \\
& \cellcolor{gray!15}Repeated & \cellcolor{gray!15}0.37 & \cellcolor{gray!15}97.32 & \cellcolor{gray!15}6.89 \\
& Phrases & 0.12 & 99.26 & 6.93 \\
& \cellcolor{gray!15}Patterns & \cellcolor{gray!15}0.48 & \cellcolor{gray!15}99.27 & \cellcolor{gray!15}7.02 \\
& All & 0.98 & 98.37 & 5.99 \\
\Xhline{1pt}
\end{tabular}
\end{table}

\subsection{RQ3: Detection Mechanism Validation}
BackFlush detection leverages backdoor susceptibility amplification: adding backdoors to an already backdoored model induces lower loss compared to a clean model. This principle parallels Membership Inference Attacks (MIA), where adversaries classify samples by analyzing loss distributions. Here, we classify models as backdoored or clean using the same approach.

To verify this, we partition $\mathcal{D}_{\textit{probe}}$ into $\mathcal{D}_1$ and $\mathcal{D}_2$. We train $\mathcal{M}_s$ with 9 backdoors on $\mathcal{D}_1$. Then, both $\mathcal{M}_s$ and clean $\mathcal{M}_{\textit{clean}}$ are fine-tuned with one additional trigger on $\mathcal{D}_2$ for 400 steps using Llama-3.2-1B, and their loss values are compared.

\begin{figure}[htbp]
    \centering
    \includegraphics[width=0.45\linewidth]{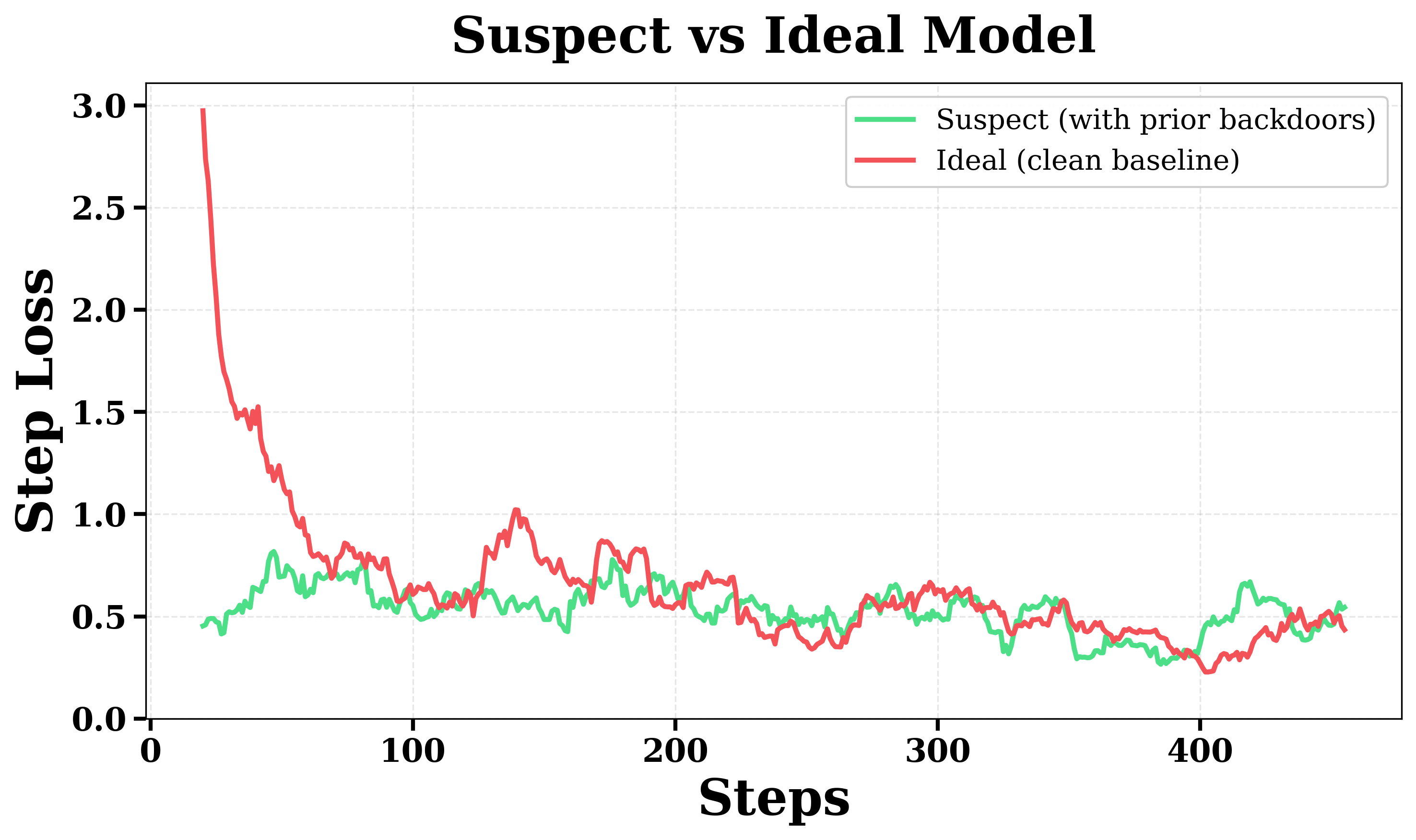}
    \hfill
    \includegraphics[width=0.45\linewidth]{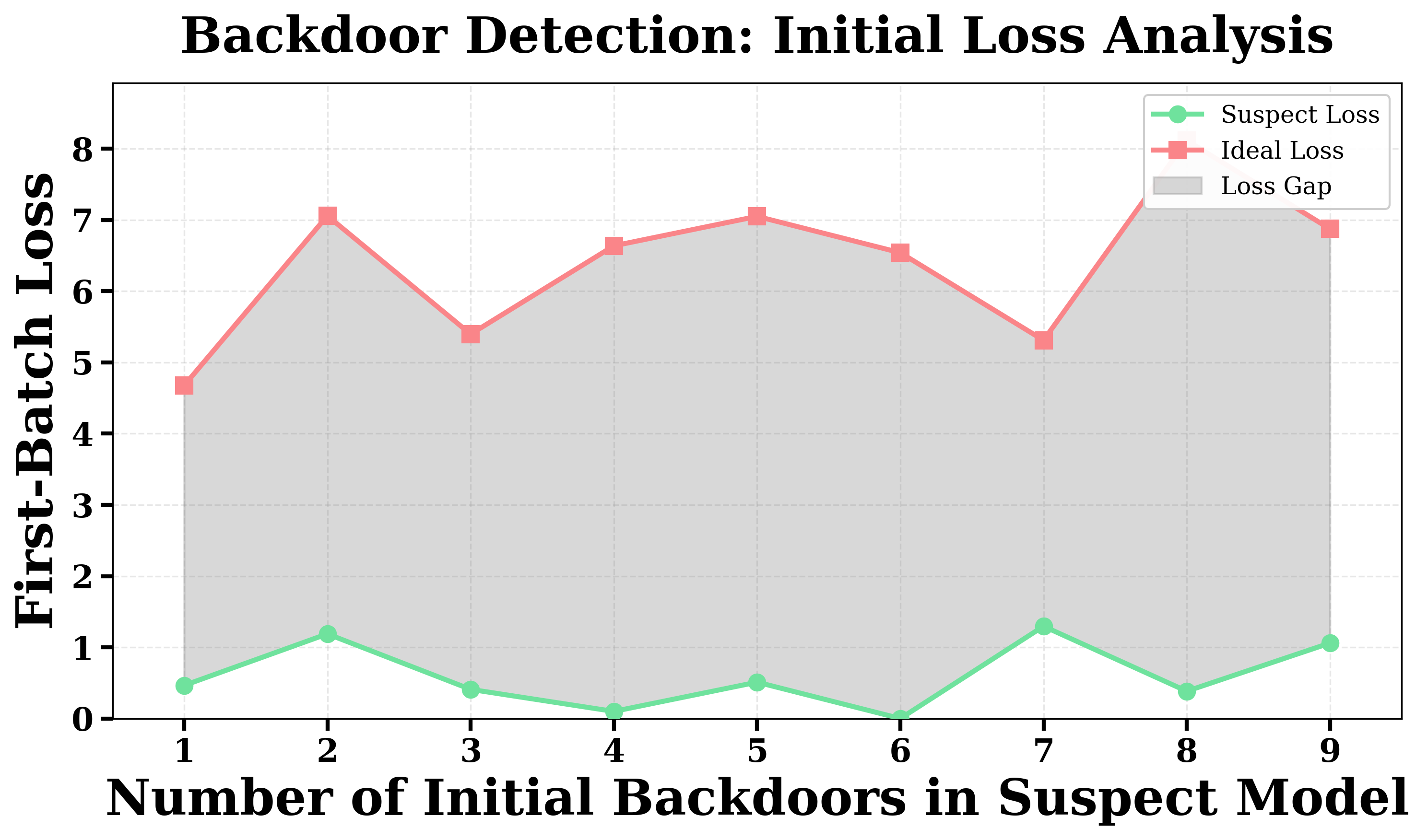}
    
    \vspace{0.5em}
    \makebox[0.45\linewidth]{\footnotesize (a) Loss comparison}
    \hfill
    \makebox[0.45\linewidth]{\footnotesize (b) Detection gap analysis}
    \vspace{0.8em}
    \caption{BackFlush detection on Llama-3.2-1B: (a) loss curves showing lower initial loss for backdoored models, (b) consistent detection gap independent of initial backdoor count.}
    \label{fig:backdoor_detection}
\end{figure}

Figure~\ref{fig:backdoor_detection}(a) shows the loss comparison over the first 20 training steps. The backdoored model exhibits significantly lower initial loss when adding new triggers, confirming that backdoor insertion is easier in already compromised models. Despite the initial loss gap, the heavily parameterized model converges quickly, with both losses becoming similar within fewer steps—making initial loss observation critical for detection.

One might question whether the initial loss gap merely reflects imbalanced backdoor counts (9 vs. 0)? Figure~\ref{fig:backdoor_detection}(b) investigates detection effectiveness as $\mathcal{M}_s$ contains varying numbers of backdoors (1 to 9+). The gap remains consistent regardless of initial backdoor count, demonstrating BackFlush's robust detection capability.

\subsection{RQ4: RoPE Unlearning Dynamics}
\begin{figure}[H]
    \centering
    \includegraphics[width=0.48\linewidth]{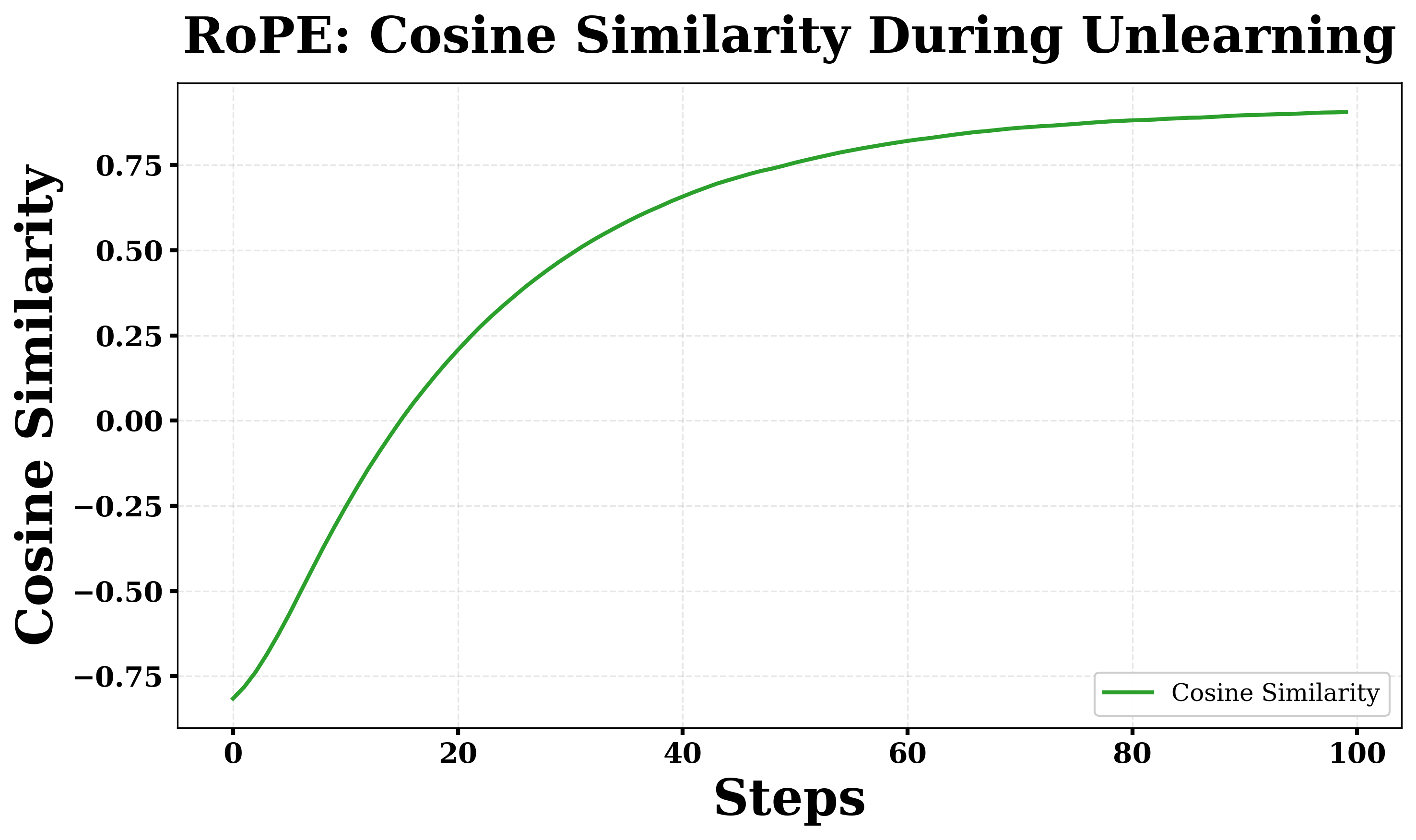}
    \hfill
    \includegraphics[width=0.48\linewidth]{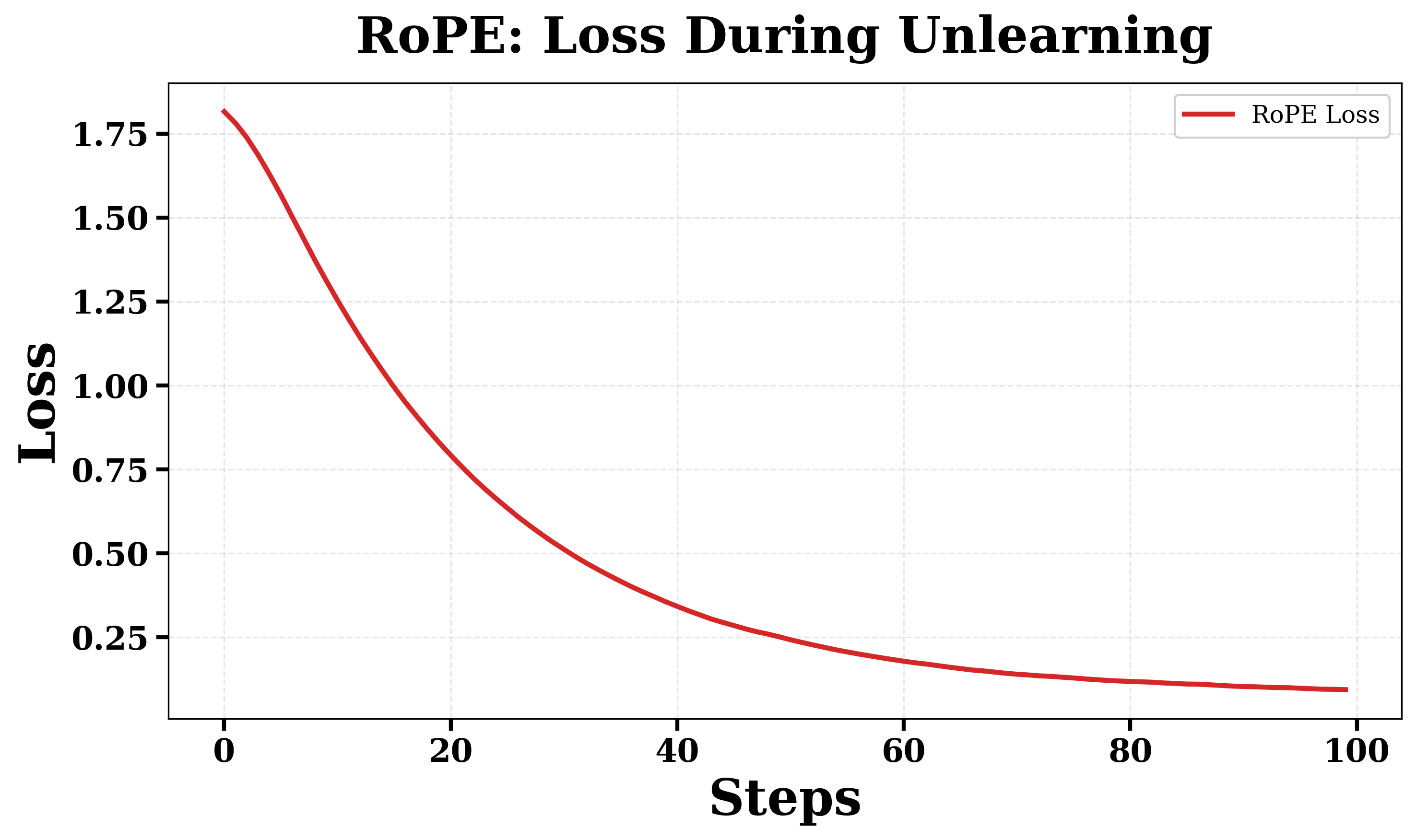}
    
    \vspace{0.3em}
    \makebox[0.48\linewidth]{\small (a) Cosine similarity:}
    \hfill
    \makebox[0.48\linewidth]{\small (b) Loss: $\approx 2$ to $\approx 0$}
    
    \caption{RoPE unlearning dynamics showing (a) cosine similarity progression and (b) loss convergence.}
    \label{fig:rope_analysis}
\end{figure}

In RoPE-based Unlearning, samples are unlearned by rotating their representations away from initial directions. Figure~\ref{fig:rope_analysis} illustrates the unlearning dynamics. Figure~\ref{fig:rope_analysis}(a) shows the cosine similarity between target and current directions, starting from $\approx -1$ and reaching $\approx +1$, indicating successful rotation to the target direction. Figure~\ref{fig:rope_analysis}(b) shows the loss curve dropping from $\approx 2$ to $\approx 0$ within 100 batches, demonstrating effective unlearning convergence.

\section{Conclusion}
To conclude, we introduce BackFlush, the first framework addressing backdoor elimination while preserving watermarks in LLMs. Our work establishes two novel observations: \textit{Backdoor Flushing Phenomenon}, demonstrating that auxiliary data injection and removal eliminates pre-existing backdoors, and \textit{Backdoor Susceptibility Amplification}, enabling constant-time detection independent of vocabulary size. The proposed BackFlush variants achieving the first watermark-preserving backdoor defense. Empirical results demonstrate $\approx 1\%$ ASR, $\approx 99\%$  CACC, and preserved watermarks across diverse trigger types, capabilities no existing method simultaneously provides. BackFlush-RoPE maintains model utility while BackFlush-GA degrades utility despite effective removal, validating RoPE unlearning's superiority. Future work extends BackFlush to backdoor removal in image generation tasks.

\bibliographystyle{IEEEtran}
\bibliography{main}

@inproceedings{zeng2024beear,
  title={Beear: Embedding-based adversarial removal of safety backdoors in instruction-tuned language models},
  author={Zeng, Yi and Sun, Weiyu and Huynh, Tran and Song, Dawn and Li, Bo and Jia, Ruoxi},
  booktitle={Proceedings of the 2024 Conference on Empirical Methods in Natural Language Processing},
  pages={13189--13215},
  year={2024}
}

@inproceedings{li2025simulate,
  title={Simulate and eliminate: Revoke backdoors for generative large language models},
  author={Li, Haoran and Chen, Yulin and Zheng, Zihao and Hu, Qi and Chan, Chunkit and Liu, Heshan and Song, Yangqiu},
  booktitle={Proceedings of the AAAI Conference on Artificial Intelligence},
  volume={39},
  number={1},
  pages={397--405},
  year={2025}
}

@article{zhang2022fine,
  title={Fine-mixing: Mitigating backdoors in fine-tuned language models},
  author={Zhang, Zhiyuan and Lyu, Lingjuan and Ma, Xingjun and Wang, Chenguang and Sun, Xu},
  journal={arXiv preprint arXiv:2210.09545},
  year={2022}
}

@article{arora2024here,
  title={Here's a free lunch: Sanitizing backdoored models with model merge},
  author={Arora, Ansh and He, Xuanli and Mozes, Maximilian and Swain, Srinibas and Dras, Mark and Xu, Qiongkai},
  journal={arXiv preprint arXiv:2402.19334},
  year={2024}
}

@article{lin2025backdoor,
  title={Backdoor Collapse: Eliminating Unknown Threats via Known Backdoor Aggregation in Language Models},
  author={Lin, Liang and Yu, Miao and Aloqaily, Moayad and Zhou, Zhenhong and Wang, Kun and Pang, Linsey and Mehrotra, Prakhar and Wen, Qingsong},
  journal={arXiv preprint arXiv:2510.10265},
  year={2025}
}

@inproceedings{zhao2025unlearning,
  title={Unlearning backdoor attacks for llms with weak-to-strong knowledge distillation},
  author={Zhao, Shuai and Wu, Xiaobao and Nguyen, Cong-Duy T and Jia, Yanhao and Jia, Meihuizi and Yichao, Feng and Tuan, Luu Anh},
  booktitle={Findings of the Association for Computational Linguistics: ACL 2025},
  pages={4937--4952},
  year={2025}
}

@article{chen2024neutralizing,
  title={Neutralizing Backdoors through Information Conflicts for Large Language Models},
  author={Chen, Chen and Sun, Yuchen and Gong, Xueluan and Gao, Jiaxin and Lam, Kwok-Yan},
  journal={arXiv preprint arXiv:2411.18280},
  year={2024}
}

@article{wang2025confguard,
  title={ConfGuard: A Simple and Effective Backdoor Detection for Large Language Models},
  author={Wang, Zihan and Zhang, Rui and Li, Hongwei and Fan, Wenshu and Jiang, Wenbo and Zhao, Qingchuan and Xu, Guowen},
  journal={arXiv preprint arXiv:2508.01365},
  year={2025}
}

@article{seddik2025pots,
  title={PoTS: Proof-of-Training-Steps for Backdoor Detection in Large Language Models},
  author={Seddik, Issam and Souihi, Sami and Tamaazousti, Mohamed and Piergiovanni, Sara Tucci},
  journal={arXiv preprint arXiv:2510.15106},
  year={2025}
}

@article{kumar2024large,
  title={Large language models (LLMs): survey, technical frameworks, and future challenges},
  author={Kumar, Pranjal},
  journal={Artificial Intelligence Review},
  volume={57},
  number={10},
  pages={260},
  year={2024},
  publisher={Springer}
}

@article{zhang2025enj,
  title={Enj: Optimizing noise with genetic algorithms to jailbreak lsms},
  author={Zhang, Yibo and Lin, Liang},
  journal={arXiv preprint arXiv:2509.11128},
  year={2025}
}

@article{yi2025safer,
  title={SaFeR-VLM: Toward Safety-aware Fine-grained Reasoning in Multimodal Models},
  author={Yi, Huahui and Wang, Kun and Li, Qiankun and Yu, Miao and Lin, Liang and Xi, Gongli and Wu, Hao and Hu, Xuming and Li, Kang and Liu, Yang},
  journal={arXiv preprint arXiv:2510.06871},
  year={2025}
}

@article{wang2025comprehensive,
  title={A comprehensive survey in llm (-agent) full stack safety: Data, training and deployment},
  author={Wang, Kun and Zhang, Guibin and Zhou, Zhenhong and Wu, Jiahao and Yu, Miao and Zhao, Shiqian and Yin, Chenlong and Fu, Jinhu and Yan, Yibo and Luo, Hanjun and others},
  journal={arXiv preprint arXiv:2504.15585},
  year={2025}
}

@inproceedings{bowen2025scaling,
  title={Scaling trends for data poisoning in LLMs},
  author={Bowen, Dillon and Murphy, Brendan and Cai, Will and Khachaturov, David and Gleave, Adam and Pelrine, Kellin},
  booktitle={Proceedings of the AAAI Conference on Artificial Intelligence},
  volume={39},
  number={26},
  pages={27206--27214},
  year={2025}
}

@article{fu2024poisonbench,
  title={Poisonbench: Assessing large language model vulnerability to data poisoning},
  author={Fu, Tingchen and Sharma, Mrinank and Torr, Philip and Cohen, Shay B and Krueger, David and Barez, Fazl},
  journal={arXiv preprint arXiv:2410.08811},
  year={2024}
}

@article{wang2024badagent,
  title={Badagent: Inserting and activating backdoor attacks in llm agents},
  author={Wang, Yifei and Xue, Dizhan and Zhang, Shengjie and Qian, Shengsheng},
  journal={arXiv preprint arXiv:2406.03007},
  year={2024}
}

@article{baumgartner2024best,
  title={Best-of-venom: Attacking rlhf by injecting poisoned preference data},
  author={Baumg{\"a}rtner, Tim and Gao, Yang and Alon, Dana and Metzler, Donald},
  journal={arXiv preprint arXiv:2404.05530},
  year={2024}
}

@article{gu2019badnets,
  title={Badnets: Evaluating backdooring attacks on deep neural networks},
  author={Gu, Tianyu and Liu, Kang and Dolan-Gavitt, Brendan and Garg, Siddharth},
  journal={Ieee Access},
  volume={7},
  pages={47230--47244},
  year={2019},
  publisher={IEEE}
}

@inproceedings{dong2024survey,
  title={A survey on in-context learning},
  author={Dong, Qingxiu and Li, Lei and Dai, Damai and Zheng, Ce and Ma, Jingyuan and Li, Rui and Xia, Heming and Xu, Jingjing and Wu, Zhiyong and Chang, Baobao and others},
  booktitle={Proceedings of the 2024 conference on empirical methods in natural language processing},
  pages={1107--1128},
  year={2024}
}

@inproceedings{huang2023semicvt,
  title={Semicvt: Semi-supervised convolutional vision transformer for semantic segmentation},
  author={Huang, Huimin and Xie, Shiao and Lin, Lanfen and Tong, Ruofeng and Chen, Yen-Wei and Li, Yuexiang and Wang, Hong and Huang, Yawen and Zheng, Yefeng},
  booktitle={Proceedings of the IEEE/CVF Conference on Computer Vision and Pattern Recognition},
  pages={11340--11349},
  year={2023}
}

@article{hubinger2024sleeper,
  title={Sleeper agents: Training deceptive llms that persist through safety training},
  author={Hubinger, Evan and Denison, Carson and Mu, Jesse and Lambert, Mike and Tong, Meg and MacDiarmid, Monte and Lanham, Tamera and Ziegler, Daniel M and Maxwell, Tim and Cheng, Newton and others},
  journal={arXiv preprint arXiv:2401.05566},
  year={2024}
}

@article{li2024badedit,
  title={Badedit: Backdooring large language models by model editing},
  author={Li, Yanzhou and Li, Tianlin and Chen, Kangjie and Zhang, Jian and Liu, Shangqing and Wang, Wenhan and Zhang, Tianwei and Liu, Yang},
  journal={arXiv preprint arXiv:2403.13355},
  year={2024}
}

@article{qiu2024megen,
  title={Megen: Generative backdoor in large language models via model editing},
  author={Qiu, Jiyang and Ma, Xinbei and Zhang, Zhuosheng and Zhao, Hai},
  journal={arXiv preprint arXiv:2408.10722},
  year={2024}
}

@article{kong2025wolf,
  title={Wolf Hidden in Sheep's Conversations: Toward Harmless Data-Based Backdoor Attacks for Jailbreaking Large Language Models},
  author={Kong, Jiawei and Fang, Hao and Yang, Xiaochen and Gao, Kuofeng and Chen, Bin and Xia, Shu-Tao and Wang, Yaowei and Zhang, Min},
  journal={arXiv preprint arXiv:2505.17601},
  year={2025}
}

@misc{qwen2.5,
    title = {Qwen2.5: A Party of Foundation Models},
    url = {https://qwenlm.github.io/blog/qwen2.5/},
    author = {Qwen Team},
    month = {September},
    year = {2024}
}

@article{grattafiori2024llama,
  title={The llama 3 herd of models},
  author={Grattafiori, Aaron and Dubey, Abhimanyu and Jauhri, Abhinav and Pandey, Abhinav and Kadian, Abhishek and Al-Dahle, Ahmad and Letman, Aiesha and Mathur, Akhil and Schelten, Alan and Vaughan, Alex and others},
  journal={arXiv preprint arXiv:2407.21783},
  year={2024}
}

@misc{jiang2023mistral7b,
      title={Mistral 7B}, 
      author={Albert Q. Jiang and Alexandre Sablayrolles and Arthur Mensch and Chris Bamford and Devendra Singh Chaplot and Diego de las Casas and Florian Bressand and Gianna Lengyel and Guillaume Lample and Lucile Saulnier and Lélio Renard Lavaud and Marie-Anne Lachaux and Pierre Stock and Teven Le Scao and Thibaut Lavril and Thomas Wang and Timothée Lacroix and William El Sayed},
      year={2023},
      eprint={2310.06825},
      archivePrefix={arXiv},
      primaryClass={cs.CL},
      url={https://arxiv.org/abs/2310.06825}, 
}

@article{welbl2017crowdsourcing,
  title={Crowdsourcing multiple choice science questions},
  author={Welbl, Johannes and Liu, Nelson F and Gardner, Matt},
  journal={arXiv preprint arXiv:1707.06209},
  year={2017}
}

@article{eldan2023tinystories,
  title={Tinystories: How small can language models be and still speak coherent english?},
  author={Eldan, Ronen and Li, Yuanzhi},
  journal={arXiv preprint arXiv:2305.07759},
  year={2023}
}

@article{joshi2017triviaqa,
  title={Triviaqa: A large scale distantly supervised challenge dataset for reading comprehension},
  author={Joshi, Mandar and Choi, Eunsol and Weld, Daniel S and Zettlemoyer, Luke},
  journal={arXiv preprint arXiv:1705.03551},
  year={2017}
}

@inproceedings{zhang2024remark,
  title={$\{$REMARK-LLM$\}$: A robust and efficient watermarking framework for generative large language models},
  author={Zhang, Ruisi and Hussain, Shehzeen Samarah and Neekhara, Paarth and Koushanfar, Farinaz},
  booktitle={33rd USENIX Security Symposium (USENIX Security 24)},
  pages={1813--1830},
  year={2024}
}

@article{kirchenbauer2023reliability,
  title={On the reliability of watermarks for large language models},
  author={Kirchenbauer, John and Geiping, Jonas and Wen, Yuxin and Shu, Manli and Saifullah, Khalid and Kong, Kezhi and Fernando, Kasun and Saha, Aniruddha and Goldblum, Micah and Goldstein, Tom},
  journal={arXiv preprint arXiv:2306.04634},
  year={2023}
}

@article{he2022cater,
  title={Cater: Intellectual property protection on text generation apis via conditional watermarks},
  author={He, Xuanli and Xu, Qiongkai and Zeng, Yi and Lyu, Lingjuan and Wu, Fangzhao and Li, Jiwei and Jia, Ruoxi},
  journal={Advances in Neural Information Processing Systems},
  volume={35},
  pages={5431--5445},
  year={2022}
}

@article{li2024can,
  title={Can multiple-choice questions really be useful in detecting the abilities of LLMs?},
  author={Li, Wangyue and Li, Liangzhi and Xiang, Tong and Liu, Xiao and Deng, Wei and Garcia, Noa},
  journal={arXiv preprint arXiv:2403.17752},
  year={2024}
}

@article{gao2023effectiveness,
  title={On the effectiveness of adversarial training against backdoor attacks},
  author={Gao, Yinghua and Wu, Dongxian and Zhang, Jingfeng and Gan, Guanhao and Xia, Shu-Tao and Niu, Gang and Sugiyama, Masashi},
  journal={IEEE Transactions on Neural Networks and Learning Systems},
  volume={35},
  number={10},
  pages={14878--14888},
  year={2023},
  publisher={IEEE}
}

@inproceedings{chen2025putting,
  title={Putting People in LLMs’ Shoes: Generating Better Answers via Question Rewriter},
  author={Chen, Junhao and Wang, Bowen and Jiang, Zhouqiang and Nakashima, Yuta},
  booktitle={Proceedings of the AAAI Conference on Artificial Intelligence},
  volume={39},
  number={22},
  pages={23577--23585},
  year={2025}
}

@article{huang2023look,
  title={Look before you leap: An exploratory study of uncertainty measurement for large language models},
  author={Huang, Yuheng and Song, Jiayang and Wang, Zhijie and Zhao, Shengming and Chen, Huaming and Juefei-Xu, Felix and Ma, Lei},
  journal={arXiv preprint arXiv:2307.10236},
  year={2023}
}

@misc{rachapudi2026repairinteractivemachineunlearning,
      title={RePAIR: Interactive Machine Unlearning through Prompt-Aware Model Repair}, 
      author={Jagadeesh Rachapudi and Pranav Singh and Ritali Vatsi and Praful Hambarde and Amit Shukla},
      year={2026},
      eprint={2604.12820},
      archivePrefix={arXiv},
      primaryClass={cs.AI},
      url={https://arxiv.org/abs/2604.12820}, 
}

@misc{rachapudi2026bidloraparameterefficientframeworkcontinual,
      title={BID-LoRA: A Parameter-Efficient Framework for Continual Learning and Unlearning}, 
      author={Jagadeesh Rachapudi and Ritali Vatsi and Praful Hambarde and Amit Shukla},
      year={2026},
      eprint={2604.12686},
      archivePrefix={arXiv},
      primaryClass={cs.LG},
      url={https://arxiv.org/abs/2604.12686}, 
}

@InProceedings{Mishra_2026_WACV,
    author    = {Mishra, Ujjwal and Shukla, Vinita and Hambarde, Praful and Shukla, Amit},
    title     = {Improvise, Adapt, Overcome -- Telescopic Adapters for Efficient Fine-tuning of Vision Language Models in Medical Imaging},
    booktitle = {Proceedings of the IEEE/CVF Winter Conference on Applications of Computer Vision (WACV)},
    month     = {March},
    year      = {2026},
    pages     = {7605-7615}
}

\end{document}